\documentclass[showpacs,aps,prb,superscriptaddress,twocolumn]{revtex4-1}

	\usepackage{textcomp} 
	\usepackage{amsmath}
	\usepackage{amssymb}
	\usepackage{graphicx} 
	\usepackage{ulem}
	\usepackage[breaklinks=true,colorlinks,citecolor=blue,linkcolor=blue,urlcolor=blue]{hyperref}
	\usepackage{mathtools}
	\usepackage{bm}
	\usepackage{color}
	\usepackage{textcomp}

	\newcommand{\ket}[1]{|{#1}\rangle}

	\definecolor{sapphire}{rgb}{0.03, 0.03, 0.41}

\begin{document}
\title{Absorption refrigerators based on Coulomb-coupled single-electron systems}
	
\author{Paolo Andrea Erdman}
\affiliation{NEST, Scuola Normale Superiore and Istituto Nanoscienze-CNR, I-56127 Pisa, Italy}
\email{paolo.erdman@sns.it}

\author{Bibek Bhandari}
\affiliation{NEST, Scuola Normale Superiore and Istituto Nanoscienze-CNR, I-56127 Pisa, Italy}

\author{Rosario Fazio}
\affiliation{ICTP, Strada Costiera 11, I-34151 Trieste, Italy}
\affiliation{NEST, Scuola Normale Superiore and Istituto Nanoscienze-CNR, I-56127 Pisa, Italy}

\author{Jukka P. Pekola}
\affiliation{QTF Centre of Excellence, Department of Applied Physics, Aalto University School of Science, P.O. Box 13500, 00076 Aalto, Finland}

\author{Fabio Taddei}
\affiliation{NEST, Istituto Nanoscienze-CNR and Scuola Normale Superiore, I-56126 Pisa, Italy}

\begin{abstract}
We analyze a simple implementation of  an absorption refrigerator, a system that requires heat and not work to achieve refrigeration, based on two Coulomb coupled single-electron systems. We analytically determine the general condition to achieve cooling-by-heating, and we determine the system parameters that simultaneously maximize the cooling power and cooling coefficient of performance (COP) finding that the system displays a particularly simple COP that can reach Carnot's upper limit. We also find that the cooling power can be indirectly determined by measuring a charge current. Analyzing the system as an autonomous Maxwell demon, we find that the highest efficiencies for information creation and consumption can be achieved, and we relate the COP to these efficiencies. Finally, we propose two possible experimental setups based on quantum dots or metallic islands that implement the non-trivial cooling condition. Using realistic parameters, we show that these systems, which resemble existing experimental setups, can develop an observable cooling power. 
\end{abstract}

\pacs{72.20.Pa,73.23.-b}

\maketitle

\section{Introduction}
Absorption refrigerators, also known in literature as self-contained or autonomous refrigerators, are systems that extract heat from a cold thermal bath only by exploiting the incoherent interaction with other two thermal baths held at higher temperatures.
No work is provided to the system, i. e. {\it cooling is achieved by heating}. The exploration for solid state implementations of absorption refrigerators has been recently attracting a considerable attention~\cite{pekola2007,linden2010,skrzypczyk2011,brunner2012,cleuren2012,kosloff2014,brunner2014,correaPalao2014, correa2014,correaPre2014, klimovsky2014, brask2015,mitchison2015, silva2015,doyeux2016,silva2016,he2017,marchegiani2017}.
The question of identifying the smallest absorption quantum refrigerators was addressed by Linden et al. in Ref.~\onlinecite{linden2010}, where systems such as two qubits, a qubit and a qutrit, or a single qutrit were considered.
It has been later shown that  these ``minimal'' systems can operate at Carnot efficiency\cite{skrzypczyk2011,brunner2012}, and the role of quantum coherence and entanglement has been addressed~\cite{brunner2014,correaPalao2014,klimovsky2014,brask2015,mitchison2015,doyeux2016}.
Besides being of fundamental interest in quantum thermodynamics, absorption refrigeration is also appealing for practical reasons: waste heat can be used to achieve cooling at the nanoscale without providing work nor requiring any external control of the system. 
There are already few experimental proposals\cite{chen2012,cleuren2012,mari2012,venturelli2013,leggio2015,hofer2016,mitchison2016,benenti2017,sanchez2017njp},
but the only experimental realization so far has been performed with trapped ions \cite{maslennikov2018}. 
In Ref.~\onlinecite{benenti2017}, in particular, it was pointed out that the very simple setup consisting of two capacitively-coupled quantum dots
could act as an absorption refrigerator, and the conditions under which its coefficient of performance (COP) can reach Carnot's limit were discussed (no entanglement or quantum coherence is required).

In this paper, on one hand, we analyze in detail a setup consisting of two capacitively-coupled quantum dots.
More precisely, we derive the general conditions under which the system operates as an absorption refrigerator, and determine the optimal system parameters which simultaneously maximize the cooling power and the COP.
We find that, under these conditions, the system exhibits a particularly simple refrigeration COP, which can indeed reach Carnot's upper limit, and that the cooling power is directly proportional to a measurable charge current~\cite{benenti2017}, allowing for an indirect measurement of a heat flow [notice that heat currents can be also measured directly in metallic islands (MIs), e.~g. in Ref.~\onlinecite{koski2015}].
Furthermore, we analyze the system as an autonomous Maxwell demon~\cite{strasberg2013, horowitz2014, koski2014prl, zhang2015, koski2016, kutvonen2016, kutvonen_pre2016, averin2017}, finding that it can operate attaining the highest efficiencies for information creation and consumption, and determining the expression that relate its COP to these efficiencies.
Finally, we propose two experimental realizations, based either on quantum dots (QDs) or metallic islands, which can implement the non-trivial requirements for the system to behave as an absorption refrigerator. We demonstrate that these systems, which closely resemble existing experimental setups ~\cite{mcclure2007,shinkai2009,shinkai2009prl,bischoff2015,hartmann2015, koski2015,thierschmann2015,volk2015,keller2016,singh2017}, can attain an observable cooling power using realistic parameters. 
\begin{figure}[!htb]
	\centering
	\includegraphics[width=1\columnwidth]{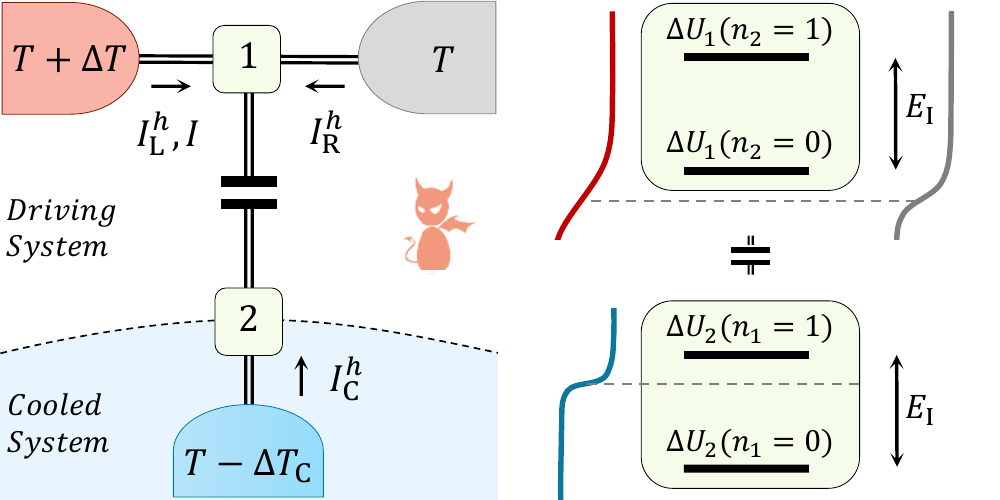}
	(a)\hspace{0.45\columnwidth} (b)
	\vspace{0.03\columnwidth}
	
	\includegraphics[width=1\columnwidth]{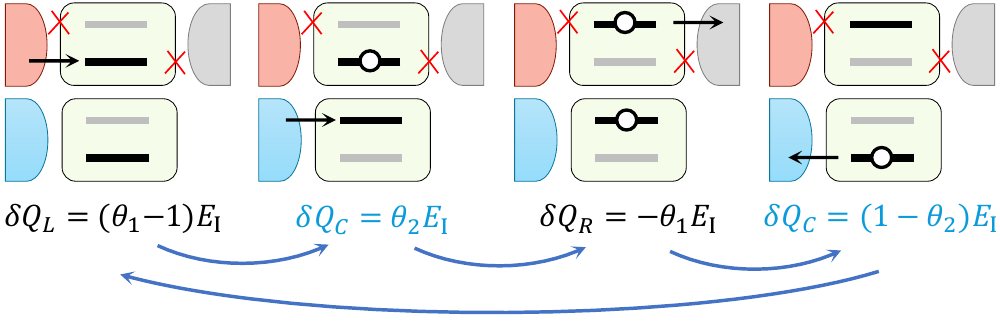}
	(c)	
	\caption{Panel (a): schematic representation of the system. Panel (b): the Fermi distribution of the leads (red upper left, gray upper right and blue lower left) is shown vertically. The black thick lines represent the transition energies $\Delta U_1(n_2)$ and $\Delta U_2(n_1)$ [Eq.~(\ref{eq:du_teta})] that are measured with respect to common chemical potential of the leads (black dashed line). Panel (c): sequence of system states and electron transitions that provide cooling when conditions (\ref{eq:gl1in_cond}) and (\ref{eq:gr0out_cond}), represented by the red crosses, are satisfied. The black horizontal lines represent the actual transition energies as determined by the occupation of the other QD, while the grey horizontal lines represent the transition energies when the other QD has opposite occupation. $\delta Q_\alpha$, for $\alpha$=L,R,C, represents the heat extracted from reservoir $\alpha$ during the corresponding electron transition.}
	\label{fig:configuration}
\end{figure}

\section{Ideal setup}
The system under investigation, depicted in Fig.~\ref{fig:configuration}(a), consists of two electronic reservoirs [upper left (L) and upper right (R)] tunnel coupled to a QD, denoted by $1$.
A second QD, $2$, capacitively coupled to $1$, is tunnel coupled to a third electronic reservoir (C).
The number of electrons occupying each Coulomb-blockaded QD can be controlled through a gate of capacitance $C_{{\rm g}i}$ and applied voltage $V_{{\rm g}i}$, with $i=1,2$.
Reservoir L is kept at a higher temperature, $T_\text{L} = T+\Delta T$, with respect to the other reservoirs which are kept at temperature $T_\text{R} = T$ and $T_\text{C}= T-\Delta T_\text{C}$. The heat current leaving reservoir $\alpha=\text{L,R,C}$ is denoted by $I^h_\alpha$, and the charge current flowing between reservoirs L and R is denoted by $I$.
We describe the transport in the entire system using a master equation approach in the sequential tunneling limit. Although we expect higher order tunneling processes, such as co-tunneling, to decrease the cooling power, these corrections are suppressed if the conductances of the junctions are much smaller than the conductance quantum and temperature is not too small. The electrostatic energy of the system is given by
\begin{multline}
	U(n_1,n_2) = E_{C1}(n_1 - n_{x1})^2 + E_{C2} (n_2-n_{x2})^2  \\
+ E_{\text I} (n_1-n_{x1})(n_2-n_{x2}) ,
\label{eq:electrostatic}
\end{multline}
where $n_i$ (for $i=1,2$) is the number of electrons in QD $i$, $n_{xi}=V_{{\rm g}i}C_{{\rm g}i}/e$, and $E_{Ci} = e^2/(2C_i)$ is its charging energy.
$C_i$ is the capacitance of QD $i$ to its surroundings, and $E_{\text I}$ is the inter-system charging energy which is controlled by the capacitive coupling between the QDs. By assuming that $E_{Ci} \gg k_BT$ and constraining the values of $n_{xi}$ to an appropriate range, we can restrict our analysis to 4 charge states, described by $n_1,n_2 = 0,1$. The ``transition energy'', i.e. the energy necessary to add an electron to QD $1$ ($2$), which also depends on the occupation of QD $2$ ($1$), is given by $\Delta U_1(n_2) = U(1,n_2) - U(0,n_2)$ [$\Delta U_2(n_1) = U(n_1,1) - U(n_1,0)$].
Since $\Delta U_i(1) - \Delta U_i(0) = E_{\text I}$, we can write
\begin{align}
\Delta U_i(n) =\theta_i E_{\text I} +(n-1)E_{\text I} ,
\label{eq:du_teta}
\end{align}
where 
\begin{align}
&\theta_1=1-n_{x2} +\frac{E_{C1}}{E_{\text I}}(1-2n_{x1}),  \\
&\theta_2=1-n_{x1} +\frac{E_{C2}}{E_{\text I}}(1-2n_{x2}),
\end{align}
can be varied using the gate voltages. The transition energies are schematically represented in Fig.~\ref{fig:configuration}(b) and \ref{fig:configuration}(c) as black thick lines.
Let $\Gamma_{\text{L/R}}^{\text{in}}(n_2)$ [$\Gamma_{\text{L}/\text{R}}^{\text{out}}(n_2)$] be the rate of electrons tunneling from (to) reservoir L/R to (from) QD $1$, and let $\Gamma_{\text{C}}^{\text{in}}(n_1)$ [$\Gamma_{\text{C}}^{\text{out}}(n_1)$] be the rate of electrons tunneling from (to) reservoir C to (from) QD $2$.
Note that the tunneling rates satisfy the detailed balance conditions 
\begin{align}
\Gamma_\alpha^{\text{out}}(n) = \exp{\left[\dfrac{\delta_\alpha(n)}{ k_BT_\alpha}\right]} \Gamma_\alpha^{\text{in}}(n),
\label{eq:dbe}
\end{align}
where $\delta_\text{L}(n)=\delta_\text{R}(n)=\Delta U_1(n)$ and $\delta_\text{C}(n)=\Delta U_2(n)$.
The currents can be calculated by specifying the tunneling rates for each process and by determining the probability $P_{n_1,n_2}$ for the two QDs to have occupation numbers $n_1$ and $n_2$ (see App.~\ref{app:master_eq}). We also use Eq.~(\ref{eq:dbe}) to express $\Gamma_\alpha^{\text{in}}(0)$ in terms of $\Gamma_\alpha^{\text{out}}(0)$ and $\Gamma_\alpha^{\text{out}}(1)$ in terms of $\Gamma_\alpha^{\text{in}}(1)$. We emphasize, however, that the results we present in the next section do not depend on the specific form of the rates, as long as Eq.~(\ref{eq:dbe}) is satisfied. Only a quantitative description of the cooling power will explicitly depend on the rates.

\section{Optimal Rates for Cooling Power and COP}
\label{sec:cooling}
The COP for refrigeration is defined as
\begin{equation}
	\eta = \frac{I^h_\text{C}}{I^h_\text{L}},
	\label{eq:eta_def}
\end{equation}
where $I^h_\text{L}>0$ is the input heat and $I^h_\text{C}>0$, the cooling power, is the heat extracted from reservoir C (their expressions are reported in App.~\ref{app:master_eq}).
Considering generic rates that are only constrained by satisfying the detailed balance condition [Eq.~(\ref{eq:dbe})], we find that the cooling power is maximized, at fixed values of $E_{\text I}$, $\theta_1$ and $\theta_2$, when
\begin{align}
	&\Gamma_\text{L}^{\text{in}}(1)=0,
	\label{eq:gl1in_cond} \\
	&\Gamma_\text{R}^{\text{out}}(0) =0,
	\label{eq:gr0out_cond} 
\end{align}
and $\Gamma_\text{L}^{\text{out}}(0)$, $\Gamma_\text{R}^{\text{in}}(1)$, $\Gamma_\text{C}^{\text{out}}(0)$, $\Gamma_\text{C}^{\text{in}}(1)$,  are as large as possible (see App.~\ref{app:optimal_rates} for details).
In this situation [i.~e. when Eqs.~(\ref{eq:gl1in_cond}) and (\ref{eq:gr0out_cond}) hold and when $\theta_i>1/2$, see App.~\ref{app:master_eq} for details] the condition for the positivity of $I^h_\text{C}$ reduces to the simple inequality
\begin{equation}
	\theta_1 > \theta_1^* \equiv 1 + \frac{1}{\eta_\text{C}^\text{h}\eta_\text{C}^\text{r}},
	\label{eq:cool_cond_ul}
\end{equation}
where $\eta_\text{C}^\text{h}= 1 - T/T_\text{L}$ and $\eta_\text{C}^\text{r}=T_\text{C}/(T-T_\text{C})$. Remarkably, in this situation the COP is also maximized (at least for $\Delta T_\text{C}=0$), and takes a particularly simple (i.~e. independent of temperatures) form
\begin{equation}
	\eta = \frac{1}{\theta_1 - 1} ,
	\label{eq:univ_eff}
\end{equation}
that only depends on $\theta_1$ (which is determined by both gate voltages $V_{\text{g1}}$ and $V_{\text{g2}}$).
Note that Eq.~(\ref{eq:cool_cond_ul}) implies that $\Delta U_1(1)>0$ and $\Delta U_1(0)>0$, i.~e. both transition energies are above the common chemical potential of the reservoirs\footnote{}, as shown in Fig.~\ref{fig:configuration}(b). This observation holds also for generic rates that do not satisfy Eqs.~(\ref{eq:gl1in_cond}) and (\ref{eq:gr0out_cond}), see App.~\ref{app:optimal_rates} for details.

Eq.~(\ref{eq:univ_eff}) implies that the input heat is always smaller than the cooling power for $\theta_1<2$, and $\eta$ is a decreasing function of $\theta_1$. The COP $\eta$ takes its maximum value when $\theta_1 = \theta_1^*$ [see Eq.~(\ref{eq:cool_cond_ul})], the smallest value of $\theta_1$ for which the system behaves as a refrigerator, giving
\begin{equation}
	\eta_{\text{max}} \equiv \eta_\text{C}^\text{h}\eta_\text{C}^\text{r},
	\label{eq:eta_max}
\end{equation}
as expected for absorption refrigerators.
Indeed, Eq.~(\ref{eq:eta_max}) states that $\eta_{\text{max}}$ can be interpreted as the combination of two two-terminal reversible machines each operating at Carnot's efficiency.
The first one is a reversible Carnot heat engine that produces work by using the temperature difference between reservoirs L and R, with $\eta_\text{C}^\text{h}= 1 - T/T_\text{L}$, while the second one is a reversible Carnot refrigerator operating between reservoirs C and R that is powered by the work of the heat engine, with $\eta_\text{C}^\text{r}=T_\text{C}/(T-T_\text{C})$. $\eta_{\text{max}}$ is the highest COP allowed by the second principle of thermodynamics, as can be proven by imposing energy conservation and zero entropy production, which read
\begin{align}
	&I^h_{\text{L}} + I^h_{\text{R}} + I^h_{\text{C}} = 0,
	\label{eq:energy_conserv} \\
	&\frac{I^h_{\text{L}}}{T_{\text{L}}} + \frac{I^h_{\text{R}}}{T_{\text{R}}} + \frac{I^h_{\text{C}}}{T_{\text{C}}} = 0.
\end{align}
Finally, when the COP is given by Eq.~(\ref{eq:eta_max}), we find that the cooling power vanishes. 

Another remarkable consequence of conditions (\ref{eq:gl1in_cond}) and (\ref{eq:gr0out_cond}), also noted in Refs.~\onlinecite{sanchez2011,benenti2017}, is that
\begin{equation}
	I^h_\text{C}=\frac{E_{\text I}}{e} I.
	\label{eq_ic_ih}
\end{equation}
Since the coupling between the upper and lower systems $E_{\text I}$ is a measurable system parameter, Eq.~(\ref{eq_ic_ih}) allows an indirect measurement of the cooling power simply by measuring the charge current in the upper system. 

A simple picture of these results can be given using the energy scheme of Fig.~\ref{fig:configuration}(b) and the conditions (\ref{eq:gl1in_cond}) and (\ref{eq:gr0out_cond}) [represented by red crosses in Fig.~\ref{fig:configuration}(c)].
The sequence of electron transitions that leads to the removal of heat from reservoir C is shown in Fig. \ref{fig:configuration}(c) and represented by blue arrows.
For each step the heat exchanged in the corresponding transition is indicated as $\delta Q_\alpha$ (for example, in the first step $\delta Q_{\rm L}=\Delta U_1(0)=(\theta_1-1)E_{\text I}$ is the input heat provided by L and associated to an electron tunneling from L to QD $1$, event which can only occur when QD $2$ is unoccupied).
\begin{figure}[!b]
	\centering
	\includegraphics[width=1\columnwidth]{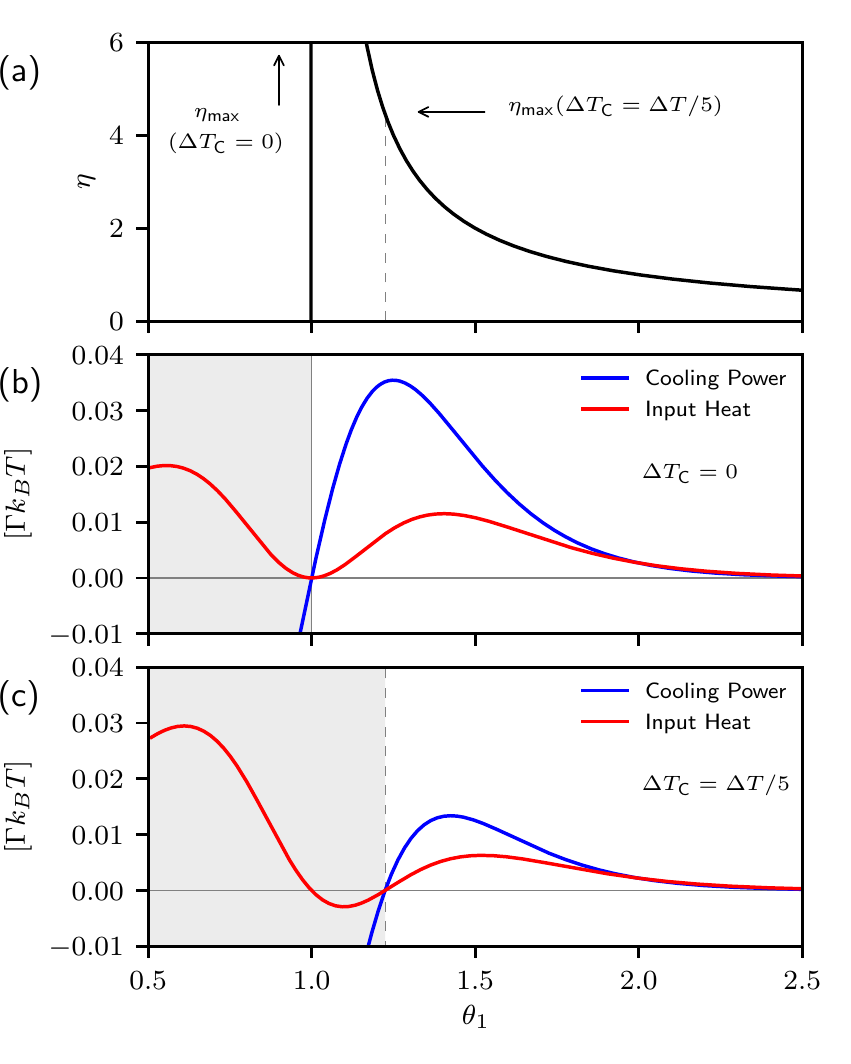}
	\caption{The coefficient of performance, COP, [panel (a)] and the heat currents in units of $\Gamma k_BT$ [panels (b) and (c)] are plotted as a function of $\theta_1$, when Eqs.~(\ref{eq:gl1in_cond}) and (\ref{eq:gr0out_cond}) are satisfied. Panel (b) refers to $\Delta T_\text{C}=0$, while panel (c) refers to $\Delta T_\text{C}= \Delta T/5$. The parameters are $\Gamma_\text{C}^{\text{out}}(0)=\Gamma_\text{C}^{\text{in}}(1)= \Gamma_\text{L}^{\text{out}}(0) = \Gamma_\text{R}^{\text{in}}(1) \equiv \Gamma$, $\theta_2 = 1$, $E_{\text{I}} = 6k_BT$ and $\Delta T/T = 1/10$. Since all rates are proportional to $\Gamma$, the heat currents depend linearly on the rate, so the plots in panel (b) and (c) do not depend on the value of $\Gamma$. }
	\label{fig:power_eff}
\end{figure}
In one cycle, an electron is transferred from L to R, and an amount $\delta Q_C^{\text{tot}} = E_{\text I}$ of heat is extracted from C: this statement is equivalent to Eq.~(\ref{eq_ic_ih}). Moreover, we notice that an amount $\delta Q_L^{\text{tot}} = (\theta_1-1)E_{\text I}$ of input heat is provided by L.
Computing the COP over one cycle as $\delta Q^{\text{tot}}_\text{C}/\delta Q^{\text{tot}}_\text{L}$ yields precisely Eq.~(\ref{eq:univ_eff}).
Eqs.~(\ref{eq:gl1in_cond}) and (\ref{eq:gr0out_cond}) guarantee that the system can only evolve along the cycle represented in blue arrows in Fig. \ref{fig:configuration}(c), or in the opposite direction, which leads to heating of reservoir C. Cooling is obtained when the system evolution along the blue arrows prevails over the opposite direction, and this happens when Eq.~(\ref{eq:cool_cond_ul}) is satisfied.

In Fig.~\ref{fig:power_eff} we plot the cooling power $I^h_{\rm C}$ and input heat $I^h_{\rm L}$, as functions of $\theta_1$, for the case $\Delta T_\text{C}=0$ [panel (b)] and $\Delta T_\text{C}= \Delta T/5$ [panel (c)] by imposing that Eqs.~(\ref{eq:gl1in_cond}) and (\ref{eq:gr0out_cond}) are satisfied.
The COP, given by a particularly simple law [Eq.~(\ref{eq:univ_eff})], is plotted in Fig.~\ref{fig:power_eff}(a).
The grey region in Fig.~\ref{fig:power_eff}(b) and \ref{fig:power_eff}(c) denotes the values of $\theta_1$ where the system does not act as a refrigerator for reservoir C [according to Eq.~(\ref{eq:cool_cond_ul}), $\theta_1^*=1$ for $\Delta T_{\rm C}=0$ and $\theta_1^*\simeq 1.2$ for $\Delta T_{\rm C}=\Delta T/5$ and $\Delta T/T=1/10$].
Fig.~\ref{fig:power_eff}(b) shows that the cooling power is zero when $\theta_1=\theta^*_1=1$ [where the COP diverges, see panel (a)] and it is maximum when $\theta_1\simeq 1.2$, where $\eta \approx 5$ [see panel (a)].
Fig.~\ref{fig:power_eff}(c), relative to $\Delta T_C = \Delta T/5$, shows that both the maximum cooling power and the corresponding COP decrease, with respect to the $\Delta T_\text{C}=0$ case, since we are refrigerating a colder system. The value of the cooling power weakly depends on $\theta_2$ in the range between $0$ and $1$.

\section{Experimental Proposals}
\label{sec:exp}
The experimental realization of the proposed absorption refrigerator relies on the ability of implementing the crucial conditions (\ref{eq:gl1in_cond}) and (\ref{eq:gr0out_cond}).
Such conditions could be, in principle, implemented by properly engineering the tunneling barrier which couple QD $1$ to its reservoirs, in order to obtain tunneling rates for QD $1$ that depend on the occupation of QD $2$.
In this section, we make use of an additional QD \cite{sanchez2011} to implement the crucial condition (\ref{eq:gl1in_cond}) that is found to be sufficient for obtaining heat extraction.

\begin{figure}[htb!]
\includegraphics[width=0.49\textwidth]{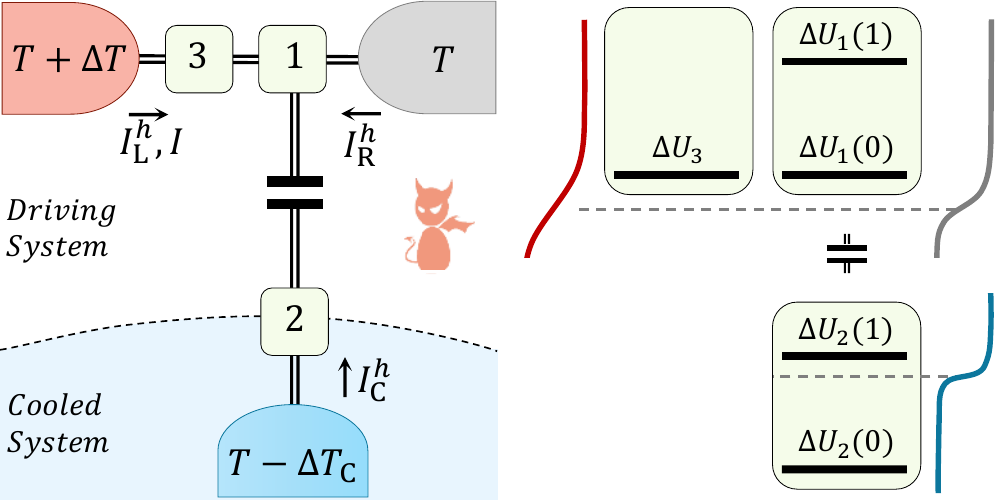}
\caption{Left: schematic representation of the system, where 1, 2 and 3 represents either QDs or MIs. Right: representation of the transition energies in the case of the system with QDs. See Fig.~\ref{fig:configuration} for details.}
\label{fig:exp_scheme}
\end{figure}

In the setup, schematically pictured in Fig.~\ref{fig:exp_scheme}, we introduce an additional QD ($3$), tunnel-coupled to $1$, and we require that its transition energy $\Delta U_3$ is aligned with $\Delta U_1(0)$ [see Fig.~\ref{fig:exp_scheme}(b)].
This way, the ``energy filtering'' effect of QD $3$ is used to suppress $\Gamma_\text{L}^{\text{in}}(1)$ with respect to $\Gamma_\text{L}^{\text{out}}(0)$.
To perform a quantitative analysis, we study the dynamics of the system of the three QDs altogether under the assumption that the coupling between QDs $1$ and $3$ is much weaker than the coupling between such QDs are their reservoirs.
The electrostatic energy of the system [see Eq.~(\ref{eq:electrostatic}) for two QDs] now takes the form
\begin{multline}
U(n_1,n_2,n_3) = E_{C1}(n_1 - n_{x1})^2 + E_{C2} (n_2-n_{x2})^2  \\
+E_\text{C3}(n_3-n_{x3})^{2}+ E_{\text I} (n_1-n_{x1})(n_2-n_{x2}) ,
\label{u3}
\end{multline}
where we have added the third term, relative to the additional QD ($3$).
Analogously to the two-QD case, we define $\Delta U_1(n_2)=U(1,n_2,n_3)-U(0,n_2,n_3)$, $\Delta U_2(n_1)=U(n_1,1,n_3)-U(n_1,0,n_3)$ and $\Delta U_3=U(n_1,n_2,1)-U(n_1,n_2,0)$, which can be written as
\begin{equation}
\begin{aligned}
	& \Delta U_1(n_2)= E_{\text I} (\theta_1 + n_2-1) \\
	& \Delta U_2(n_1) = E_{\text I} (\theta_2 + n_1-1) \\
	& \Delta U_3 = E_{\text I}(\theta_3-1) ,
\end{aligned} 
\end{equation}
where we have defined the following 3 independent dimensionless parameters
\begin{equation}
\begin{aligned}
	&\theta_1 = (1-2n_{x1}) E_{\rm C1}/E_{\text I} + (1-n_{x2}) \\
	&\theta_2 = (1-2n_{x2}) E_{\rm C2}/E_{\text I} + (1-n_{x1})\\
	&\theta_3 = (1-2n_{x3}) E_{\rm C3}/E_{\text I} + 1 .
\end{aligned} 
\end{equation}

If we assume that each QD can be only singly-occupied, we can restrict our analysis to the following 8 states: $|0,0,0\rangle$, $|0,0,1\rangle$, $|0,1,0\rangle$, $|1,0,0\rangle$, $|1,0,1\rangle$, $|0,1,1\rangle$, $|1,1,0\rangle$ and $|1,1,1\rangle$, where $|n_1,n_2,n_3\rangle$ is the state associated to  the set of occupation numbers $(n_1,n_2,n_3)$.
The probability $p_\alpha$ for the system to be in the state $|\alpha\rangle=|n_1,n_2,n_3\rangle$ is calculated by solving the master equation in the stationary case (see App.~\ref{app:master3} for details)
\begin{equation}
\dot{p}_{\alpha}=\sum_{\nu}\left(-\Gamma_{\alpha \nu}\, p_{\alpha}+\Gamma_{\nu \alpha}\, p_{\nu}\right) ,
\label{mastereqns}
\end{equation}
where $\Gamma_{\alpha \nu}$ is the rate for the transition from state $|\alpha\rangle$ to state $|\nu\rangle$.
The rates $\Gamma_{\alpha \nu}$ which account for the transfer of electrons between a QD and a reservoir can be expressed as~\cite{kaasbjerg2016}
\begin{equation}
\Gamma_{\alpha \nu}=\hbar^{-1}\gamma_{\lambda}f_{\lambda}(\Delta \tilde{U}_{\alpha\nu}) ,
\label{tunnelrate}
\end{equation}
where $\gamma_{\lambda}$ is the coupling energy between the reservoir  $\lambda=\lambda(\alpha,\nu )$ and a QD, where $\lambda=$ L, R, C depends on the initial state $|\alpha\rangle$ and final state $|\nu\rangle$.
In Eq.~(\ref{tunnelrate}), $f_{\lambda}(\epsilon)=[1+e^{\epsilon/({k_BT_{\lambda}})}]^{-1}$ is the reservoir Fermi distribution function, while $\Delta \tilde{U}_{\alpha\nu}=\tilde{U}(\nu)-\tilde{U}(\alpha)$ is the transition energy, where $\tilde{U}(\alpha)=U(n_1,n_2,n_3)$ [see Eq.~(\ref{u3})] with the set of occupation numbers corresponding to the state $|\alpha\rangle$.
The inter-dot transition rates, which account for the transfer of electrons between QD 1 and 3 [namely, $\Gamma_{(0,0,1),(1,0,0)}$ and $\Gamma_{(0,1,1),(1,1,0)}$], are obtained using the procedure outlined in App.~\ref{app:master3} under the assumption that the hopping element $t$ is much smaller than the coupling energy between QDs and reservoirs~\cite{gurvitz1998,hazelzet2001,sztenkiel2007,dong2004,dong2008}.

The relevant heat currents can now be written as
\begin{equation}
I_\text{C,L}^{h}=\sum_{\alpha \nu}\Delta \tilde{U}_{{\alpha \nu}}\left(\Gamma_{ \alpha \nu}\, p_{\alpha}-\Gamma_{\nu \alpha}\,p_{\nu}\right) ,
\label{heatcurrent}
\end{equation}
where the sum runs over the states specified in App.~\ref{app:master3}.
\begin{figure}[tb!]
	\centering
	\includegraphics[width=1\columnwidth]{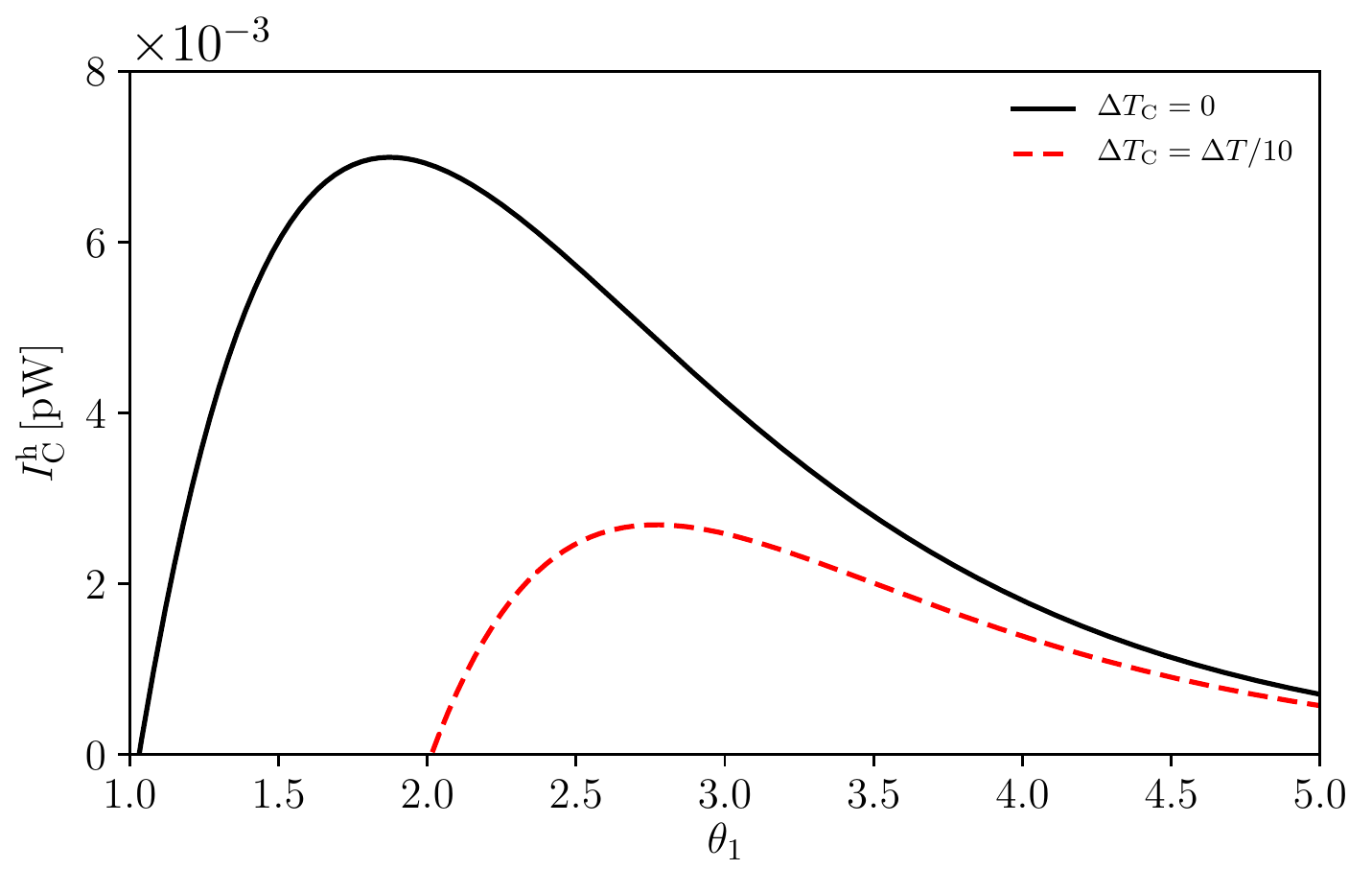}
	\caption{Cooling power $I_\text{C}^h$, relative to the system containing three QDs and represented in Fig.~\ref{fig:exp_scheme}, under resonant condition [$\Delta U_3 = \Delta U_1(0)$]. $I_\text{C}^h$ is plotted as a function of $\theta_1$ for the case $\Delta T_\text{C}=0$ (solid black curve) and the case $\Delta T_\text{C}=\Delta T/10$ (dashed red curve), setting $\theta_2=1/2$ and imposing $\theta_3=\theta_1$. The parameters are of the order of the experimental ones reported in Ref.~\onlinecite{keller2016} and read: $E_{\text I}=0.72$ meV, $\gamma_{\text{L}}=\gamma_{\text{R}}=\gamma_{\text{C}}=0.036$ meV, $t=0.016$ meV, and $T=\Delta T=4.17$ K.}
	\label{fig:three_qds}
\end{figure}
In Fig.~\ref{fig:three_qds} we plot the cooling power $I_\text{C}^h$, as a function of $\theta_1$, for realistic parameters and setting $\theta_3=\theta_1$ in order to obtain the resonant condition [i.~e. $\Delta U_3 = \Delta U_1(0)$] which approximately implements condition (\ref{eq:gl1in_cond}).
The solid black curve is relative to the case $\Delta T_\text{C}=0$, while the dashed red curve refers to $\Delta T_\text{C} = \Delta T/10$.
Fig.~\ref{fig:three_qds} shows that in both cases heat extraction is obtained and that $I_\text{C}^h$ takes a maximum value of the order of $10^{-2}$ pW.
We notice that, as in the ideal case, the cooling power is weakly dependent on $\theta_2$ in the range between 0 and 1, and that in this case $I_\text{C}^h$ is maximized for $\theta_2\simeq 1/2$.
Moreover, we check that when the difference between $\Delta U_3$ and $\Delta U_1(0)$ is not much larger than the coupling energies $\gamma_\text{L/R/C}$,  the condition $\theta_3=\theta_1$ is essentially fulfilled and the curves in Fig.~\ref{fig:three_qds} do not change appreciably.
We have demonstrated that the implementation of the crucial condition (\ref{eq:gl1in_cond}) alone is sufficient to obtain heat extraction.
Cooling power, as seen above, is expected to be maximal when the additional condition (\ref{eq:gr0out_cond}) is also satisfied. This could be implemented by adding another filtering QD in series with $1$, between R and $1$, and aligning its transition energy to $\Delta U_1(1)$. 
For experimental purposes, however, a simpler system is desirable, especially because the transition energies of the different QDs need to be tuned by individual gates (not shown in Fig.~\ref{fig:exp_scheme}), operation that is further complicated by possible cross-couplings arising between them.

\subsection{Metallic islands}
We will now explore the possibility of replacing the QDs in the setup depicted in Fig.~\ref{fig:exp_scheme} with MIs. These are systems still characterized by a large charging energy but, as opposed to QDs, they present a continuous distribution of energy levels (the level spacing is much smaller than $k_BT$) so that electrons within the island are thermalized and distributed according to the Fermi distribution. 
Due to the absence of discrete levels, the sharp ``filtering effect'' discussed above in the QD system and exploited to satisfy the crucial conditions (\ref{eq:gl1in_cond}) and (\ref{eq:gr0out_cond}) is not possible.
As we will show below, however, heat extraction can nonetheless be obtained in the setup depicted in Fig.~\ref{fig:exp_scheme}, where 1, 2 and 3 are now usual metals and reservoir R (grey element) is superconducting.
Our aim is to approximately satisfy Eq.~(\ref{eq:gl1in_cond}) by properly tuning the chemical potential of MI $3$. Conversely, by exploiting the superconducting gap of reservoir R, we aim at approximately satisfying Eq.~(\ref{eq:gr0out_cond}) in order to suppress the electron transfer with energy near $\Delta U_1(0)$.
Unlike the case with QDs, here the detailed balance condition [Eq.~(\ref{eq:dbe})] is not satisfied by the rates between islands at different temperatures.
As we shall see, however, this has only minor consequences.

The electrostatic energy of the system is equal to the one relative to the system of three QDs, Eq.~(\ref{u3}).
Also in this case we assume that each MI can only be singly-occupied so that our analysis can be restricted to the 8 states defined in the QD case.
In the sequential tunneling regime, the stationary probability $p_{\alpha}$ that the system is in the state $\alpha$ is computed by solving the master equation (\ref{mastereqns}), where, unlike in the QDs case, the rate for the transition from state $\alpha$ to state $\nu$ is given by
\begin{align}
\Gamma_{\alpha\nu}=\frac{1}{e^{2}R_{\alpha\nu}}\int d\epsilon \mathcal{N}_{\lambda}(\epsilon)\mathcal{N}_{\mu}&(\epsilon-\Delta\tilde{U}_{\nu\alpha})f_{\lambda}(\epsilon)\nonumber \\
&\left[1-f_{\mu}(\epsilon-\Delta \tilde{U}_{\nu\alpha})\right].
	\label{eq:gamma_def}
\end{align}
Here, $R_{\alpha\nu}$ is the resistance of the tunneling barrier involved in the tunneling process, while $\lambda=\lambda(\alpha,\nu)$ and $\mu=\mu(\alpha,\nu)$ identify the indices of the MIs or reservoirs involved in the tunneling process.
In Eq.~(\ref{eq:gamma_def}), $\mathcal{N}_\lambda$ denote the normalized density of states, which takes the value $\mathcal{N}_\lambda = 1$ for $\lambda=$ 1,2,3,L,C, and
\begin{equation}
	\mathcal{N}_\text{R}(\epsilon) = \left| Re\left( \frac{\epsilon+i\gamma}{\sqrt{(\epsilon+i\gamma)^2 - \Delta^2}} \right) \right| ,
\end{equation}
for the superconducting reservoir \cite{dynes1978,pekola2010}.
Here $\gamma$ is a phenomenological inverse quasi-particle lifetime, and $\Delta$ is the superconducting gap.
As before, the heat currents $I^h_\text{L}$ and $I^h_\text{C}$ are defined as the heat currents extracted from reservoirs L and C, and are computed in App.~\ref{app:heat_mi}.

\begin{figure}[!tb]
	\centering
	\includegraphics[width=1\columnwidth]{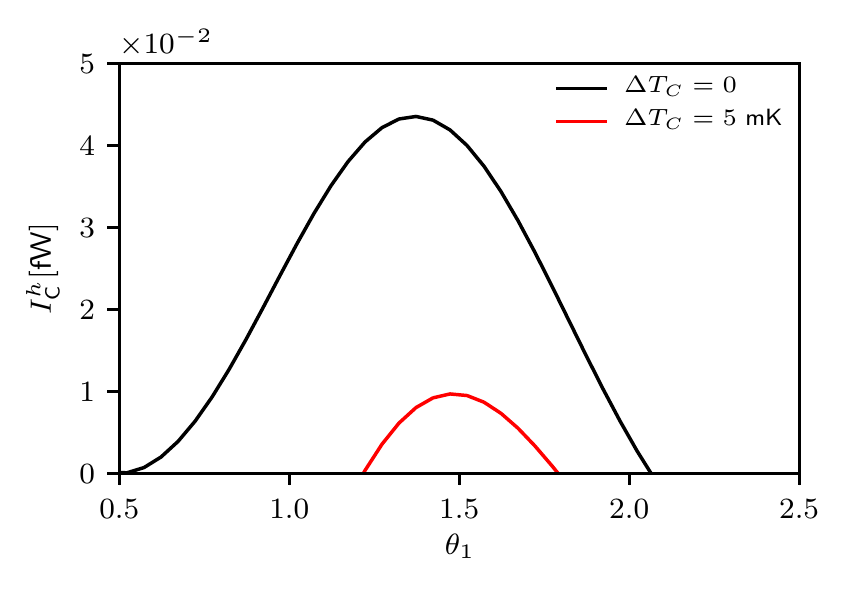}
	\caption{Cooling power, relative to the setup depicted in Fig.~\ref{fig:exp_scheme} for MIs, as a function of $\theta_1$ for two different values of $\Delta T_\text{C}$, and setting $\theta_2=1/2$ and $\theta_3 = \theta_1+1/2$. The parameters used are experimentally relevant, see for example Refs.~\onlinecite{koski2015,dutta2017}, and read: $E_{\text I}=25$ $\mu$eV, $\Delta = 35$ $\mu$eV, $\gamma = 10^{-3}$ $\mu$eV, $T = 100$ mK, $\Delta T = 200$ mK and $R_{\alpha\nu}=10$ k$\Omega$ for all barriers.}
	\label{fig:cooling}
\end{figure}

In Fig.~\ref{fig:cooling} the cooling power is plotted, using realistic parameters, as a function of $\theta_1$, for $\Delta T_\text{C}=0$ (solid black curve) and for $\Delta T_C=5$ mK (dashed red curve) and setting $\theta_2=1/2$.
We assume that MIs 1 and 3 are at temperature $T$, while MI 2 is at temperature $T-\Delta T_\text{C}$.
Aiming at implementing the condition (\ref{eq:gl1in_cond}), we place the electrochemical potential $\Delta U_3$ half way between $\Delta U_1(0)$ and $\Delta U_1(1)$, i.~e. we set $\theta_3= \theta_1+1/2$.
In fact, this guarantees that (if $k_BT\lesssim E_\text{I}$) the electron energy distribution in MI 3 is such that electron transfer to MI 1 is suppressed in the case where MI 2 is occupied.
Note, however, that the opposite process (electron transfer from 1 to 3) is not suppressed.
Indeed, to obtain heat extraction we need to further assume that electrode R is superconducting.
Figure~\ref{fig:cooling} shows that cooling is achieved in both cases, $\Delta T_\text{C}=0$ and $\Delta T_\text{C}=5$ mK.
In the former case, the maximum cooling power is of the order $10^{-2}$ fW, while in the latter heat extraction is still possible, but the maximum cooling power decreases roughly by a factor $4$.
Interestingly, heat extraction occurs even for $\theta_1<1$, contrary to the prediction of Eq.~(\ref{eq:cool_cond_ul}).
This can be attributed to the fact that the detailed balance condition (\ref{eq:dbe}) is not satisfied for the tunneling rates coupling MIs or reservoirs having different temperatures.
An amount of heat equal to $I^h_\text{C}$ is also extracted from MI $2$ (see App.~\ref{app:heat_mi} for details).
Naturally no heat is extracted when reservoir R is in the normal state.
We find that $I^h_\text{C}$ is maximized when $\theta_2\simeq 1/2$ and $\theta_3\simeq \theta_1+1/2$, and that its increase with $\Delta T$ is at most linear.
Nevertheless, we wish to point out that there is no simple condition to identify the optimal values of $E_{\text I}$ and $\Delta$.
Yet by scaling all energies and temperatures of a given factor, the cooling power scales as the square of such factor.

\section{Maxwell demon: mutual information flow}
\label{sec:demon}
Recent experimental advancements have turned the intriguing MD thought experiment~\cite{leff2014,bennett1982} into real experiments, spurring a vast experimental and theoretical research. A profound relation between information and thermodynamics was found \cite{sagawa2008,sagawa2010,esposito2012,deffner2013,barato2014,horowitz2014,parrondo2015} and various manifestation of MDs have been theoretically\cite{averin2011,mandal2012,sanchez2012,barato2013,bergli2013,strasberg2013,strasberg2014,zhang2015,kutvonen2016,
kutvonen_pre2016,tanabe2016,averin2017,sanchez2017} and experimentally\cite{serreli2007,price2008,thorn2008,raizen2009,toyabe2010,berut2012,koski2014pnas,koski2014prl,roldan2014,
koski2015,chida2015,camati2016,koski2016,thierschmann2016,vidrighin2016,chida2017} studied.  In autonomous MDs, where the demon is part of the analyzed system, cooling has been studied from various standpoints, but, as far as we know, in all cases a voltage bias was used to ``power'' the demon.
Conversely, our system does not require work, but it can be viewed as an autonomous MD since there is no direct heat transfer between the driving (D) and the cooled (C) system associated with electron tunneling; the cooling effect can thus be interpreted as due to information transfer.

According to the theoretical framework developed in Ref.~\onlinecite{horowitz2014}, one can write the following inequalities 
\begin{align}
	&\dot{\mathcal{S}}^{(r)}_\text{D} - \dot{\mathcal{I}} \geq 0, 	\label{eq:entropy_d} \\
	&\dot{\mathcal{S}}^{(r)}_\text{C} + \dot{\mathcal{I}} \geq 0,
	\label{eq:entropy}
\end{align}
where $\dot{\mathcal{S}}^{(r)}_\text{D} = -I^h_\text{L}/T_\text{L} - I^h_\text{R}/T_\text{R}$ and $ \dot{\mathcal{S}}^{(r)}_\text{C} = -I^h_\text{C}/T_\text{C}$ represent, respectively, the entropy variation in the driving and cooled reservoirs, while $\dot{\mathcal{I}}$ ($-\dot{\mathcal{I}}$) represents the variation of mutual information between system D and C due to tunneling events in D (C). The system behaves as a refrigerator, by extracting heat from  reservoir C, when $\dot{\mathcal{S}}^{(r)}_\text{C}<0$, which implies $\dot{\mathcal{I}}>0$ in order to satisfy Eq.~(\ref{eq:entropy}). We can thus interpret system D as a MD which acquires information by monitoring system C. In turn, system C uses this information as a resource to decrease its temperature. Eq.~(\ref{eq:entropy}) shows that the cooling of reservoir C is bounded by $\dot{\mathcal{S}}^{(r)}_\text{C} \geq -\dot{\mathcal{I}}$, while Eq.~(\ref{eq:entropy_d}) shows that reservoirs L and R are bound to dissipate at least $\dot{\mathcal{S}}^{(r)}_\text{D} \geq \dot{\mathcal{I}}$. This observation motivates the definition of the following thermodynamic efficiencies \cite{horowitz2014} 
\begin{align}
	&\eta_\text{D} = \frac{\dot{\mathcal{I}}}{\dot{\mathcal{S}}^{(r)}_\text{D}} \leq 1, &\eta_\text{C} = \frac{|\dot{\mathcal{S}}^{(r)}_\text{C}|}{\dot{\mathcal{I}}} \leq 1,
	\label{eq:thermo_eff_def}
\end{align}
where $\eta_\text{D}$ represents the ``information generation'' efficiency, and $\eta_\text{C}$ the ``information consumption'' efficiency.
Notice that by definition $0 \leq \eta_\text{D},\eta_\text{C} \leq 1$, and they are equal to $1$ when, respectively, Eqs.~(\ref{eq:entropy_d}) and (\ref{eq:entropy}) are strict equalities.
While $\eta$ is a quantity assigned to the entire system, $\eta_\text{D}$ and $\eta_\text{C}$ characterize the two subsystems, so that they can be viewed as a refinement to $\eta$ \cite{horowitz2014}.
By combining Eqs.~(\ref{eq:eta_def}), (\ref{eq:energy_conserv}) and (\ref{eq:thermo_eff_def}), the COP $\eta$ can be written in terms of the product $\eta_{\text{D}}\eta_{\text{C}}$ and of $\eta_\text{C}^\text{r}$ as
\begin{equation}
	\eta = \eta_{\text{max}}\frac{\eta_\text{D}\eta_\text{C}}{1+ \eta_\text{C}^\text{r}(1-\eta_\text{D}\eta_\text{C})}.
	\label{eq:eff_relation}
\end{equation} 
This is consistent with the fact that, in general, $\eta_\text{D}$ and $\eta_\text{C}$ individually provide more information than $\eta$, which is directly related only to their product $\eta_\text{D}\eta_\text{C}$.
Using Eq.~(\ref{eq:eff_relation}), we notice that $\eta=\eta_{\text{max}}$ if and only if $\eta_\text{D}=\eta_\text{C}=1$. This implies that for $\theta_1=\theta_1^*$, where the COP reaches Carnot's limit [see Eq.~(\ref{eq:eta_max})], we have that $\eta_\text{D}=\eta_\text{C}=1$.

\begin{figure}[!htb]
	\centering
	\includegraphics[width=1\columnwidth]{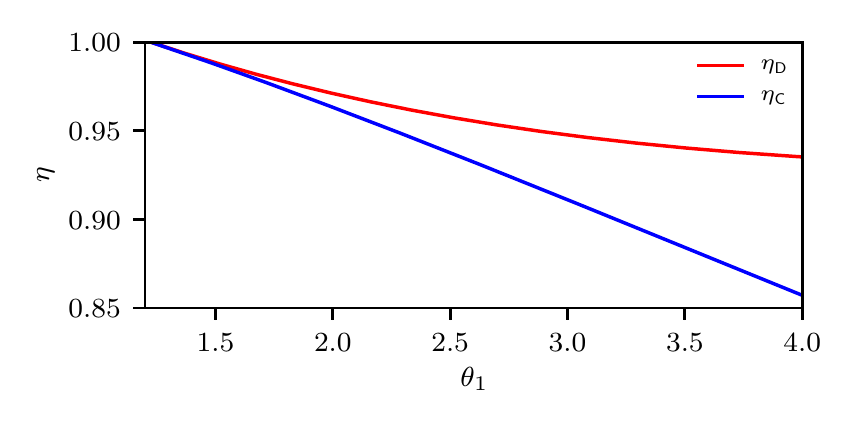}
	\caption{The efficiencies $\eta_\text{D}$ and $\eta_\text{C}$ are plotted as a function of $\theta_1$ starting from $\theta_1=\theta_1^*$ for the case $\Delta T_\text{C}= \Delta T/5$. The parameters are the same as in Fig.~\ref{fig:power_eff}(c).}
	\label{fig:thermo_eff}
\end{figure}
In Fig.~(\ref{fig:thermo_eff}) we plot $\eta_\text{D}$ and $\eta_\text{C}$ as a function of $\theta_1$.
As expected, when $\theta_1 = \theta_1^*$, $\eta_\text{D} = \eta_\text{C} =1$. For larger values of $\theta_1$ the efficiencies decrease, but they remain close to $1$. In general, finding high thermodynamic efficiencies in this model is not trivial~\cite{horowitz2014,kutvonen2016,tanabe2016,chida2017}.

\section{Conclusions}
We have studied several aspects of a minimal implementation of an absorption refrigerator based on two Coulomb coupled single-electron systems~\cite{benenti2017}.  We have derived the general condition to guarantee cooling by heating and we have found the optimal rates that simultaneously maximize cooling power and coefficient of performance (COP). A simple relation between cooling power and charge current is also found. Analyzing the system as an autonomous Maxwell demon, we have shown that the efficiencies for information production and consumption can reach their upper bounds, and we have related the COP to these efficiencies. Finally, we have put forward two experimental proposals, based on quantum dots (QDs) and metallic islands (MIs). In both proposals we have introduce an additional QD or MI that implements the non-trivial condition required to achieve cooling-by-heating. By plugging in realistic parameters we have shown that these proposals, which resemble existing experiments, yield observable heat currents~\cite{Ronzani2018,Linke2017}.

\section{Acknowledgments}
We would like to thank Robert Whitney for fruitful discussions and Gonzalo Manzano for reading the manuscript and providing useful comments. This work has been supported by SNS-WIS joint lab ``QUANTRA'', by the
SNS internal projects ``Thermoelectricity in nano-devices'', and ``Non-equilibrium dynamics of one-dimensional quantum
systems: From synchronisation to many-body localisation'', by the CNR-CONICET cooperation programme ``Energy conversion in quantum, nanoscale, hybrid devices'', by the COST ActionMP1209 ``Thermodynamics in the quantum regime'', by the Academy of Finland (grant 312057) and by the European Union's Horizon 2020 research and innovation programme under the European Research Council (ERC) programme (grant agreement 742559).

\begin{appendix}
\begin{widetext}

\section{Master Equation}
\label{app:master_eq}
The probability $P_{n_1,n_2}$ that the system is in a state with $n_1$ and $n_2$ electrons in QDs 1 and 2 is calculated by solving the following system of equations
\begin{multline}
	\begin{pmatrix}
		-\left[ \Gamma_\text{L}^{\text{in}}(0) + \Gamma_\text{R}^{\text{in}}(0) + \Gamma_\text{C}^{\text{in}}(0) \right] & \Gamma_\text{C}^{\text{out}}(0) & \Gamma_\text{L}^{\text{out}}(0) + \Gamma_\text{R}^{\text{out}}(0) & 0 \\
		0 & \Gamma_\text{L}^{\text{in}}(1) + \Gamma_\text{R}^{\text{in}}(1) & \Gamma_\text{C}^{\text{in}}(1) & -\left[ \Gamma_\text{L}^{\text{out}}(1) + \Gamma_\text{R}^{\text{out}}(1) + \Gamma_\text{C}^{\text{out}}(1) \right] \\
		\Gamma_\text{C}^{\text{in}}(0) & -\left[ \Gamma_\text{L}^{\text{in}}(1) + \Gamma_\text{R}^{\text{in}}(1) + \Gamma_\text{C}^{\text{out}}(0) \right] & 0 &  \Gamma_\text{L}^{\text{out}}(1) + \Gamma_\text{R}^{\text{out}}(1)  \\
		1 & 1 & 1 & 1 \\
	\end{pmatrix}
	\\ \times
	\begin{pmatrix}
		P_{0,0} \\ P_{0,1} \\ P_{1,0} \\ P_{1,1}
	\end{pmatrix}
	=
	\begin{pmatrix}
		0 \\ 0 \\ 0 \\ 1
	\end{pmatrix}.
	\label{eq:master_eq}
\end{multline}
The first three equations correspond to  the master equations where the time-derivatives $\dot{P}_{0,0}$, $\dot{P}_{1,1}$, and $\dot{P}_{0,1}$ are set to zero, while the last equation corresponds to the normalization requirement.
The charge current is given by
\begin{equation}
	I = e\left[ P_{0,0} \Gamma_{\text{L}}^{\text{in}}(0) + P_{0,1} \Gamma_{\text{L}}^{\text{in}}(1) - P_{1,0} \Gamma_{\text{L}}^{\text{out}}(0) - P_{1,1} \Gamma_{\text{L}}^{\text{out}}(1)  \right],
	\label{eq:i}
\end{equation}
where $e$ is the electron charge, and the heat current leaving reservoir $\alpha$ is given by
\begin{equation}
	I^h_{\alpha} =  P_{0,0} \Gamma_\alpha^{\text{in}}(0) \Delta U_1(0) - P_{1,1} \Gamma_\alpha^{\text{out}}(1) \Delta U_1(1) + P_{0,1} \Gamma_\alpha^{\text{in}}(1) \Delta U_1(1) - P_{1,0} \Gamma_\alpha^{\text{out}}(0) \Delta U_1(0) ,
\label{eq:ihlr}
\end{equation}
for $\alpha = \text{L,R}$, and
\begin{equation}
	I^h_{\text C} =  P_{0,0} \Gamma_{\text C}^{\text{in}}(0) \Delta U_2(0) - P_{1,1} \Gamma_{\text C}^{\text{out}}(1) \Delta U_2(1)   + P_{1,0} \Gamma_{\text C}^{\text{in}}(1) \Delta U_2(1) - P_{0,1} \Gamma_{\text C}^{\text{out}}(0) \Delta U_2(0) .
\label{eq:ihc}
\end{equation}
Note that one can exploit the symmetry of the transitions energies with respect to the common chemical potential when $\theta_1=\theta_2=1/2$ [see Eqs.~(\ref{eq:du_teta}) and Fig.~\ref{fig:configuration}(b)] to restrict the analysis to the range $\theta_1 \geq 1/2$, without loss of generality.
In fact, the heat currents relative to the case $\theta_i=\bar{\theta}_i < 1/2$ are equal to the ones obtained with $\theta_i =1- \bar{\theta}_i (> 1/2)$, while the charge currents relative to the case $\theta_i=\bar{\theta}_i < 1/2$ are equal in amplitude but with opposite sign with respect to the ones obtained with $\theta_i =1- \bar{\theta}_i (> 1/2)$.
This can be explicitly verified by substituting $\theta_i \to  1-\theta_i$ and $\Gamma^{\text{in/out}}_\alpha(n) \to \Gamma^{\text{out/in}}_\alpha(1-n)$ in Eqs. (\ref{eq:du_teta}), (\ref{eq:master_eq}), (\ref{eq:i}), (\ref{eq:ihlr}) and (\ref{eq:ihc}).
\end{widetext}

\section{Optimal rates for cooling power and COP}
\label{app:optimal_rates}
By substituting the probability $P_{n_1,n_2}$, solution of Eq.~(\ref{eq:master_eq}), into the expression (\ref{eq:ihc}) for $I^h_{\text C}$ and imposing the detailed balance condition (\ref{eq:dbe}), we find that $I^h_\text{C} >0$ if and only if 
\begin{equation}
\begin{aligned}
	&\Gamma_\text{L}^{\text{out}}(0)\Gamma_\text{R}^{\text{in}}(1)\left(e^{\,j\eta_\text{C}^{\text{h}} (\theta_1-1)} - 1\right) \\
	 	&-\Gamma_\text{L}^{\text{in}}(1)\left[\Gamma_\text{L}^{\text{out}}(0)\left(1- e^{-j\eta_\text{C}^{\text{h}}} \right) + \Gamma_\text{R}^{\text{out}}(0)\left( 1 - e^{-j\eta_\text{C}^{\text{h}} \theta_1 } \right)   \right]  \\
	  &-\left[ \Gamma_\text{L}^{\text{out}}(0) + \Gamma_\text{R}^{\text{out}}(0)  \right]\left[\Gamma_\text{L}^{\text{in}}(1) + \Gamma_\text{R}^{\text{in}}(1) \right] (e^{\,j/\eta_\text{C}^{\text{r}}}-1) >0 .
\end{aligned}
\label{eq:cool_cond_gen}
\end{equation}
Interestingly, the condition (\ref{eq:cool_cond_gen}) does not depend on the rates $\Gamma^{(\text{in/out})}_{\rm C}$ relative to the cooled system, nor on $\theta_2$.
In Eq.~(\ref{eq:cool_cond_gen}), $\eta_\text{C}^\text{h}= 1 - T/T_\text{L}$ is the Carnot efficiency of a heat engine operating between L and R, $\eta_\text{C}^\text{r}=T_\text{C}/(T-T_\text{C})$ is the Carnot COP of a refrigerator operating between R and C, and $j=E_{\text I}/k_BT$.
Restricting to the range $\theta_1\geq1/2$ (see App.~\ref{app:master_eq} for details), the first line of Eq.~(\ref{eq:cool_cond_gen}) is the only term that can be positive, so that a necessary non-trivial condition to satisfy Eq.~(\ref{eq:cool_cond_gen}) is that $\theta_1>1$.

When Eq.~(\ref{eq:cool_cond_gen}) is satisfied, at fixed $E_{\text I}$, $\theta_1$ and $\theta_2$, we find that $I_\text{C}^h$ is a decreasing function of $\Gamma_\text{L}^{\text{in}}(1)$ and $\Gamma_\text{R}^{\text{out}}(0)$, so that the optimal choice for such parameters is
\begin{align}
	\Gamma_\text{L}^{\text{in}}(1)=\Gamma_\text{R}^{\text{out}}(0) = 0 .
	\label{00}
\end{align}
Now, assuming (\ref{00}), $I_\text{C}^h$ is an increasing function of the remaining rates  $\Gamma_\text{L}^{\text{out}}(0)$, $\Gamma_\text{R}^{\text{in}}(1)$, $\Gamma_\text{C}^{\text{out}}(0)$, $\Gamma_\text{C}^{\text{in}}(1)$, so that the optimal choice is to take them as large as possible, compatibly with the validity of the sequential tunneling picture.

\section{Derivation of the master equation for the system with  three QDs}
\label{app:master3}
The Hamiltonian of the system with three QDs can be represented as
\begin{align}
H_{\rm sys}&=\sum_{\alpha}\epsilon_{\alpha}|\alpha\rangle\langle \alpha| + \nonumber \\
&+E_{\text I}\big(|1,1,0\rangle\langle 1,1,0|+|1,1,1\rangle\langle 1,1,1|\big) + \nonumber \\
&+t\big(|1,0,0\rangle\langle 0,0,1|+|1,1,0\rangle\langle 0,1,1| +{\rm h.c.}\big),
\label{hsys}
\end{align}
where $\epsilon_{\alpha}$ is the energy of state $\ket{\alpha}$ in the absence of coupling, $t$ is the hopping element between the two tunnel coupled QDs (3 and 1), and $E_{\text I}$ represents the inter-dot charging energy between the capacitively-coupled QDs, 1 and 2.
Under the assumption that the hopping element $t$ is much smaller than the coupling energy between QDs and reservoirs, in Refs.~\onlinecite{gurvitz1998,hazelzet2001,sztenkiel2007,dong2004,dong2008} it was shown that the density matrix $\rho$ (whose components are defined as $\rho_{\alpha\beta}=\langle\alpha|\rho |\beta \rangle$) satisfies a modified Liouville equation.
In particular, the diagonal components $\rho_{\alpha\alpha}$ satisfy\cite{wegewijs1999}
\begin{align}
\dot{\rho}_{\alpha\alpha}=-i[H_\text{sys},\rho]_{\alpha\alpha}-\sum_{\nu}\Gamma_{\alpha \nu}\rho_{\alpha\alpha}+\sum_{\delta}\Gamma_{\delta\alpha}\rho_{\delta\delta} ,
\label{diagon}
\end{align}
while the off-diagonal components, resulting from coherent tunneling of electrons between QDS $3$ and $1$, satisfy
\begin{align}
\dot{\rho}_{\alpha\beta}=-i[H_\text{sys},\rho]_{\alpha\beta}-\frac{1}{2}\sum_{\nu}\left(\Gamma_{\alpha\nu}+\Gamma_{\beta \nu}\right)\rho_{\alpha\beta} .
\label{offdiagon}
\end{align}
In Eqs.~(\ref{diagon}) and (\ref{offdiagon}), the first (Liouville) term contains the system Hamiltonian (\ref{hsys}), while the other terms describe the coupling of the QDs with the reservoirs.
In Eq.~(\ref{offdiagon}), $|\alpha\rangle=|0,0,1\rangle$ and $|\beta\rangle=|1,0,0\rangle$ (and viceversa), or $|\alpha\rangle=|0,1,1\rangle$ and $|\beta\rangle=|1,1,0\rangle$ (and viceversa), since the only non-zero off-diagonal terms are the ones related to electron tunneling between QDs $1$ and $3$ (with $2$ either occupied or unoccupied).
Note that Eqs.~(\ref{diagon}) and (\ref{offdiagon}) depend explicitly only on the transition rate $\Gamma_{\alpha \nu}$, from state $|\alpha\rangle$ to state $|\nu\rangle$, which accounts for the transfer of electrons between a QD and the corresponding reservoir $\lambda=\lambda(\alpha,\nu )$.
In particular, the transition rates for tunnelling events between $1$ and $3$, such as $\Gamma_{(0,0,1),(1,0,0)}$ and $\Gamma_{(0,1,1),(1,1,0)}$, do not appear in Eqs.~(\ref{diagon}) and (\ref{offdiagon}).
The rates appearing in Eqs.~(\ref{diagon}) and (\ref{offdiagon}) can be expressed as~\cite{kaasbjerg2016}
\begin{equation}
\Gamma_{\alpha \nu}=\hbar^{-1}\gamma_{\lambda}f_{\lambda}(\Delta \tilde{U}_{\alpha\nu}) ,
\end{equation}
where $\gamma_{\lambda}$ is the coupling energy between reservoir $\lambda$ and QD, $f_{\lambda}(\epsilon)=[1+e^{\epsilon/({k_BT_{\lambda}})}]^{-1}$ is the reservoir Fermi distribution function, while $\Delta \tilde{U}_{\alpha\nu}=\tilde{U}(\nu)-\tilde{U}(\alpha)$ is the transition energy, where $\tilde{U}(\alpha)=U(n_1,n_2,n_3)$ [see Eq.~(\ref{u3})] with the set of occupation numbers corresponding to the state $|\alpha\rangle$.

In order to keep the notation compact, we assign an index to each set of occupation numbers as follows: $(0,0,0)\rightarrow 0$, $(1,0,0)\rightarrow 1$, $(0,1,0)\rightarrow 2$, $(0,0,1)\rightarrow 3$, $(1,1,0)\rightarrow 4$, $(0,1,1)\rightarrow 5$, $(1,0,1)\rightarrow 6$ and $(1,1,1)\rightarrow 7$.
We will show now that the inter-dot tunneling rates, i.~e. $\Gamma_{3,1}\equiv\Gamma_{(0,0,1),(1,0,0)}$ and $\Gamma_{5,4}\equiv\Gamma_{(0,1,1),(1,1,0)}$, can be obtained by using Eqs.~(\ref{diagon}) and (\ref{offdiagon}).~\cite{sprekeler2004}
Let us consider the component $(3,3)$ of Eq.~({\ref{diagon}}), i.~e.
\begin{align}
\dot{\rho}_{3,3}=&-it\big(\rho_{1,3}-\rho_{3,1}\big)-\left(\Gamma_{3,0}+\Gamma_{3,5}+\Gamma_{3,6}\right)\rho_{3,3}\nonumber \\
&+\Gamma_{0,3}\,\rho_{0,0}+\Gamma_{5,3}\,\rho_{5,5}+\Gamma_{6,3}\,\rho_{6,6} .
\label{den33}
\end{align}
In the steady state ($\dot\rho=0$), the components $(3,1)$ and $(5,4)$ of Eq.~(\ref{offdiagon}) can be written, respectively, as
\begin{equation}
\rho_{3,1}=\frac{t\left(\rho_{3,3}-\rho_{1,1}\right)}{\epsilon_{3}-\epsilon_{1}-i\frac{\tilde{\Gamma}^\text{(0)}}{2}} 
\label{den31}
\end{equation}
and 
\begin{equation}
\rho_{5,4}=\frac{t\left(\rho_{5,5}-\rho_{4,4}\right)}{\epsilon_{5}-\epsilon_{4}-E_{\text I}-i\frac{\tilde{\Gamma}^\text{(1)}}{2}} ,
\label{den2312}
\end{equation} 
where $\tilde{\Gamma}^\text{(0)} = \Gamma_{3, 6}+\Gamma_{3,5}+\Gamma_{3,0}+\Gamma_{1,6}+\Gamma_{1,4}+\Gamma_{1, 0}$ accounts for all the processes which lead to the decay of the states $|3\rangle$ and $|1\rangle$, and $\tilde{\Gamma}^\text{(1)} = \Gamma_{5, 3}+\Gamma_{5,2}+\Gamma_{5,7}+\Gamma_{4,1}+\Gamma_{4,2}+\Gamma_{4,7}$ accounts for all the processes which lead to the decay of the states $|0,1,1\rangle$ and $|1,1,0\rangle$.
By substituting Eq.~(\ref{den31}) into Eq.~(\ref{den33}), with $\rho_{1,3}=\rho_{3,1}^*$, the latter equation will contain only diagonal elements of the density matrix, thus representing an ordinary master equation of the form
\begin{equation}
\dot{p}_{3}=\sum_{\nu=0,5,6}\left(-\Gamma_{3 \nu}\, p_{3}+\Gamma_{\nu 3}\, p_{\nu}\right) - \Gamma_{3,1}p_3 +\Gamma_{1,3} p_1,
\label{rateeq3}
\end{equation}
where $p_\alpha=\rho_{\alpha\alpha}$ represents the probability for the state $|\alpha\rangle$.
In Eq.~(\ref{rateeq3}), the two terms (in $\Gamma_{3,1}$ and $\Gamma_{1,3}$) accounting for the transitions between states $|0,0,1\rangle$ and $|1,0,0\rangle$, when QD $2$ is unoccupied, now appear.
The associated inter-dot tunneling rate takes the form
\begin{equation}
\Gamma_{3,1}=\frac{t^{2}\tilde{\Gamma}^\text{(0)}}{(\epsilon_{3}-\epsilon_{1})^{2}+\left(\frac{\tilde{\Gamma}^\text{(0)}}{2}\right)^{2}} .
\label{tunnelrate2}
\end{equation}
Similarly, using Eq.~(\ref{den2312}) in the expression for $\dot{\rho}_{5,5}$ or $\dot{\rho}_{4,4}$, one obtains the inter-dot tunneling rate
\begin{equation}
\Gamma_{5,4}=\frac{t^{2}\tilde{\Gamma}^\text{(1)}}{\left(\epsilon_{3}-\epsilon_{1}-E_{\text I}\right)^{2}+\left(\frac{\tilde{\Gamma}^\text{(1)}}{2}\right)^{2}} 
\label{tunnelrate3}
\end{equation}
in the case where QD $2$ is occupied.
Note that both inter-dot tunneling rates have a Lorentzian profile.

The relevant heat currents can be written as
\begin{equation}
I_\text{C,L}^{h}=\sum_{\alpha, \nu}\Delta \tilde{U}_{{\alpha \nu}}\left(\Gamma_{ \alpha \nu}\, p_{\alpha}-\Gamma_{\nu \alpha}\,p_{\nu}\right) ,
\label{heatcurrentapp}
\end{equation}
where the sum runs over the indices $(\alpha,\nu)=(0,2)$, $(1,4)$, $(3,5)$, $(6,7)$ for the cooling power $I_\text{C}^{h}$, and over the values $(\alpha,\nu)=(0,3)$, $(1,6)$, $(2,5)$, $(4,7)$ for the input heat $I_\text{L}^{h}$.

\section{Heat currents in the system with metallic islands}
\label{app:heat_mi}
Since MIs presents a continuum of states, the heat exchanged in a single electron transition is not fixed by the electrostatic energy difference as in Eq.~(\ref{heatcurrent}), but it depends on the energy of the electron that is tunneling. We thus need to define the following heat rates~\cite{Bhandari2018}
\begin{equation}
\begin{aligned}
	\Gamma_{\alpha\nu}^{h,\text{out}}=\frac{1}{e^{2}R_{\alpha\nu}}\int d\epsilon\, \epsilon \, \mathcal{N}_{\lambda}(\epsilon)\mathcal{N}_{\mu}(\epsilon-\Delta\tilde{U}_{\nu\alpha})f_{\lambda}(\epsilon) \\ \left[1-f_{\mu}(\epsilon-\Delta \tilde{U}_{\nu\alpha})\right]
	\nonumber
\end{aligned}
\end{equation}
and
\begin{equation}
\begin{aligned}
	\Gamma_{\alpha\nu}^{h,\text{in}}=\frac{1}{e^{2}R_{\alpha\nu}}\int d\epsilon\, (\epsilon-\Delta\tilde{U}_{\nu\alpha}) \mathcal{N}_{\lambda}(\epsilon)\mathcal{N}_{\mu}(\epsilon-\Delta\tilde{U}_{\nu\alpha})f_{\lambda}(\epsilon)\\  \left[1-f_{\mu}(\epsilon-\Delta \tilde{U}_{\nu\alpha})\right].
		\nonumber
\end{aligned}
\end{equation}
$\Gamma_{\alpha\nu}^{h,\text{out}}$ is to the heat rate extracted from $\lambda(\alpha,\nu)$ (the reservoir or island from which the electron is tunneling) and $\Gamma_{\alpha\nu}^{h,\text{in}}$ corresponds to the heat injected into $\mu(\alpha,\nu)$ (the reservoir or island to which the electron is tunneling to) when the system undergoes a transition from $\alpha$ to $\nu$. We thus have that
\begin{equation}
I_\text{C,L}^{h}=\sum_{\alpha, \nu}\left(\Gamma_{ \alpha \nu}^{h,\text{out}}\, p_{\alpha}-\Gamma_{\nu \alpha}^{h,\text{in}}\,p_{\nu}\right) ,
\label{heatcurrent_mi}
\end{equation}
where, as in Eq.~(\ref{heatcurrentapp}), the sum runs over the values $(\alpha,\nu)=(0,2)$, $(1,4)$, $(3,5)$, $(6,7)$, for $I_\text{C}^{h}$, and over $(\alpha,\nu)=(0,3)$, $(1,6)$, $(2,5)$, $(4,7)$, for $I_\text{L}^{h}$.
The heat extracted from MI $2$ can also be computed as in Eq.~(\ref{heatcurrent_mi}) by summing over the values $(\alpha,\nu)=(2,0)$, $(4,1)$, $(5,3)$, $(7,6)$.

\end{appendix}

\end{document}